\documentclass[a4paper,11pt]{article}
\usepackage{jcappub} 
\makeatletter
\gdef\@fpheader{ }
\makeatother

\hypersetup{
	colorlinks   = true, 
	urlcolor     = blue, 
	linkcolor    = blue, 
	citecolor    = red   
}
\renewcommand{\b}[1]{{\bf #1}}

\renewcommand{\Re}{\operatorname{Re}}
\renewcommand{\Im}{\operatorname{Im}}

\title{Classicality of Stochastic Noise Away From Quasi-de~Sitter Inflation}

\author{Mahdiyar Noorbala}
\emailAdd{mnoorbala@ut.ac.ir}
\affiliation[a]{Department of Physics, University of Tehran, Iran. P.O.\ Box 14395-547}

\abstract{It is well known that a coarse-grained scalar field living on a de~Sitter (dS) background exhibits classical stochastic behavior, driven by a noise whose amplitude is set by the Hubble constant $H$.  The coarse-graining is achieved by discarding wave numbers larger than a cutoff $\sigma a H$ and demanding that $\sigma \ll 1$.  Similar results hold for quasi-dS space, where the equation of state parameter $w$ is close to $-1$.  Here we present exact expressions for the noise amplitude of a free massless field on an inflationary background with constant $w < -1/3$.  We find that a classical stochastic behavior can emerge for $-5/3 < w < -1/3$.  Furthermore, as we move away from $w = -1$ and approach $w = -1/3$, the constraint $\sigma \ll 1$ is relaxed and larger cutoffs ($\sigma \sim 1$) become feasible, too.  However, in general the amplitude of the noise depends on $\sigma$, except in the quasi-dS regime $w \approx -1$.}

\begin{document}

\maketitle

\section{Introduction}

Inflation is the most popular paradigm for primordial cosmology~\cite{Guth:1980zm, Linde:1981mu, Albrecht:1982wi}.  Vacuum fluctuations of the inflaton field source curvature perturbations that seed the structure in the universe as well as the anisotropies in the cosmic microwave background.  The study of long wavelength component of the field is most easily carried out in the stochastic approach~\cite{Vilenkin:1983xp, Vilenkin:1983xq, Linde:1986fd, Starobinsky:1986fx, Rey:1986zk, Aryal:1987vn, Starobinsky:1994bd}.  This can be done both for the inflaton field itself as well as any spectator (test) field that lives on a fixed inflationary background without affecting the dynamics of the geometry.  In this paper we focus on the latter case, but in either case, the stochastic approach is based on separating the UV/IR modes via a cutoff scale $k_\sigma = \sigma a H$, where $H$ is the Hubble parameter and $\sigma$ is a dimensionless number.  By integrating out the UV modes $k>k_\sigma$, one obtains the coarse-grained field (a.k.a., the long mode, or the IR mode) and the UV modes, as they leave the horizon as a result of the accelerating expansion, turn out to play the role of a stochastic noise as a source term in a Langevin equation.  Thus the stochastic approach is essentially an effective theory for the stochastically evolving classical IR modes.  It is shown~\cite{Starobinsky:1986fx, Sasaki:1987gy} that in order for this classical picture to hold for a massive field in dS space, one requires $\sigma$ to be small, more specifically,
\begin{equation}
\exp \left(-\frac{3H^2}{m^2} \right) \ll \sigma^2 \ll \frac{m^2}{3H^2} \ll 1.
\end{equation}
Furthermore, the resulting Langevin equation governing the coarse-grained field has a source term that is a white noise with an amplitude that is independent of the cutoff parameter $\sigma$.  The Langevin equation, or the corresponding Fokker-Planck equation, can be solved and various properties of the system can be read from the correlation functions and probability distributions \cite{Nambu:1987ef, Nambu:1988je, Kandrup:1988sc, Nakao:1988yi, Nambu:1989uf, Mollerach:1990zf, Spokoiny:1993uc, Linde:1993xx, Linde:1996hg, Starobinsky:1994bd, Kunze:2006tu, Prokopec:2007ak, Prokopec:2008gw, Tsamis:2005hd, Enqvist:2008kt, Finelli:2008zg, Finelli:2010sh, Garbrecht:2013coa, Garbrecht:2014dca, Burgess:2014eoa, Burgess:2015ajz, Boyanovsky:2015tba, Boyanovsky:2015jen, Fujita:2017lfu, Glavan:2017jye, Gorbenko:2019rza, Mirbabayi:2019qtx, Mirbabayi:2020vyt, Cohen:2021fzf, Cruces:2021iwq, Cruces:2022imf}.  There is also a vast literature on the stochastic $\delta N$ formalism where the statistics of curvature perturbations is related to that of the inflaton field and which is particularly useful in studying the large fluctuations of curvature perturbations ~\cite{Fujita:2013cna, Fujita:2014tja, Vennin:2015hra, Vennin:2016wnk, Assadullahi:2016gkk, Grain:2017dqa, Pattison:2017mbe, Pattison:2018bct, Noorbala:2018zlv, Firouzjahi:2018vet, Biagetti:2018pjj, Noorbala:2019kdd, Pattison:2019hef, Talebian:2019opf, Ezquiaga:2019ftu, Firouzjahi:2020jrj, Ando:2020fjm, Ballesteros:2020sre, Pattison:2021oen, Figueroa:2020jkf, Figueroa:2021zah, Ahmadi:2022lsm, Animali:2022otk, Talebian:2022jkb, Nassiri-Rad:2022azj, Asadi:2023flu, Tomberg:2023kli, Mishra:2023lhe}.

The above picture is conventionally derived for a field living in dS or quasi-dS background where the equation of state parameter $w$ is equal or close to $-1$, i.e., the first slow-roll parameter $\epsilon = -\dot H/H^2$ is assumed to be small.  As we show in this paper, the criterion $\sigma\ll1$ is required for classicality only when $\epsilon$ is small.  Indeed, we present a situation away from $w=-1$ (i.e., with non-small $\epsilon$) in which fairly large values of the cutoff ($\sigma\sim1$) are sufficient for classicality.

To be clear, we should mention that this is not the first work to study stochastic inflation in a non-dS background.  Indeed, the general form of the noise is known to be proportional to the power spectrum of field fluctuations, regardless of the background FLRW geometry (see, for example, ref.~\cite{Grain:2017dqa}).  However, to our knowledge, when it comes to the question of classicality, either a thorough analysis is absent or it is assumed that the geometry is quasi-dS ($\epsilon\ll1$).  On the other hand, there are also studies of stochastic inflation in the non-slow-roll regime (see, for example, refs.~\cite{Pattison:2019hef, Ballesteros:2020sre, Pattison:2021oen, Figueroa:2020jkf, Figueroa:2021zah, Ahmadi:2022lsm}).  However, these are models of ultra-slow-roll~\cite{Tsamis:2003px} where the first slow-roll parameter $\epsilon$ is small and it is the second slow-roll parameter $d\log\epsilon/dN$ that is large.\footnote{As usual, we reserve the term ``slow-roll'' for the situation where all of the slow-roll parameters are small ($\forall n:\epsilon_n\ll1$, where $\epsilon_1 = -\dot H/H^2$ and $\epsilon_{n+1} = d\log\epsilon_n/dN$).  We use the term ``quasi-dS'' when the first slow-roll parameter $\epsilon_1$ is small, regardless of the smallness or largeness of the higher slow-roll parameters.  The term ``non-slow-roll'' usually means a quasi-dS regime where slow-roll is violated, but to avoid confusion,  we don't use it hereafter.  We do not work with higher slow-roll parameters either, and so we drop the index: everywhere that $\epsilon$ appears it refers to $-\dot H/H^2$.}

We demonstrate our point in a very simple setup: a free massless field on an FLRW background with arbitrary $w$ that is constant in time but not necessarily close to $-1$.  We also require inflation to take place, so we impose $w<-1/3$.  Then we investigate the condition of classicality and find that, although near $w=-1$ we need $\sigma\ll1$, as we get close to $w=-1/3$, this constraint is relaxed and $\sigma$ doesn't have to be small.

The rest of this paper is organized as follows: We review the general theory of stochastic inflation for a spectator field on a generic inflationary background in section~\ref{sec:rev}.  Then we revisit the conventional case of a free field on dS space in section~\ref{sec:dS}.  Section~\ref{sec:w} is the main part of the paper, where we present our results about the classicality criterion away from the quasi-dS regime.  Finally we summarize and conclude in section~\ref{sec:summary}.

\section{Review of Stochastic Inflation and the Criterion of Classicality}\label{sec:rev}

Let us review the derivation of the Langevin equation in stochastic inflation and see how the noise term emerges.  Along the way, we pay careful attention to the criterion of classicality.

We consider a non-interacting spectator scalar field on a fixed inflationary background $ds^2 = -dt^2 + a^2 d{\bf x}^2 = a^2 [-d\tau^2 + d{\bf x}^2]$, where $\tau$ is the conformal time.  In the Heisenberg picture, the field operator $\hat\chi(\tau,\b x)$ satisfies 
\begin{equation}\label{chi-eom-free}
\hat\chi'' + 2{\cal H}\hat\chi' + m^2 a^2 \hat\chi - \nabla^2\hat\chi = 0,
\end{equation}
where the conformal Hubble parameter ${\cal H} =a'/a$ is related to the usual Hubble parameter $H = \dot a/a$ by ${\cal H}=aH$.\footnote{We denote time derivative with respect to $t$ by a dot, and derivative with respect to $\tau$ by a prime.}  It follows that the Fourier mode $\hat\chi_{\b k}(\tau) = \int \frac{d^3x}{(2\pi)^{3/2}} \hat\chi(\tau,\b x) e^{-i\b k \cdot \b x}$ is given in terms of creation and annihilation operators by
\begin{equation}
\hat\chi_{\b k}(\tau) = \chi_k(\tau) \hat a_{\bf k} + \chi_k^*(\tau) \hat a_{-\bf k}^\dag,
\end{equation}
where the mode function $\chi_{\b k}$ is related to the Mukhanov-Sasaki variable by $u_k = a \chi_k$ which satisfies
\begin{equation}\label{Mukhanov-Sasaki}
u_k'' + \left( k^2 + m^2a^2 - \frac{a''}{a} \right) u_k = 0.
\end{equation}
We employ the standard Bunch-Davies state by imposing the asymptotic condition $u_k \to e^{-ik\tau}/\sqrt{2k}$ when the mode is deep inside the horizon ($k\gg\cal H$).

It should be noted that, although these expressions are written for generic $a(\tau)$, it must in fact be an inflationary background in the past (namely, $\ddot a>0$, or equivalently, $\epsilon < 1$), so that the modes \textit{exit} the horizon rather than \textit{enter} the horizon.  Otherwise the asymptotic boundary condition cannot be imposed in the far past.

The key idea of stochastic formalism is the short/long (UV/IR) mode decomposition with respect to a momentum cutoff $k_\sigma(\tau) = \sigma {\cal H}(\tau)$ that corresponds to a wavelength larger than horizon by the factor $\sigma^{-1}$, i.e., 
\begin{align}
\hat\chi_s(\tau,\b x) &= \int \frac{d^3k}{(2\pi)^{3/2}} \hat\chi_{\bf k}(\tau) e^{i \b k \cdot \b x} \theta \left( k - k_\sigma(\tau) \right), \\
\hat\chi_l(\tau,\b x) &= \int \frac{d^3k}{(2\pi)^{3/2}} \hat\chi_{\bf k}(\tau) e^{i \b k \cdot \b x} \theta \left( k_\sigma(\tau) - k \right),
\end{align}
where $\hat\chi_l$ is the long mode component of $\hat\chi$, $\hat\chi_s$ is its short mode component, and $\theta$ is the Heaviside step function that is employed as our window function.  Similar to the filed $\hat\chi$, the field velocity (with respect to the $e$-folding time $N$) $\hat v=d\hat\chi/dN$ can be split into the long mode $\hat v_l$ and the short mode $\hat v_s$ as follows
\begin{align}
\hat v_s(\tau,\b x) &= \int \frac{d^3k}{(2\pi)^{3/2}} \frac{d\hat\chi_{\bf k}}{dN} e^{i \b k \cdot \b x} \theta \left( k - k_\sigma(\tau) \right), \\
\hat v_l(\tau,\b x) &= \int \frac{d^3k}{(2\pi)^{3/2}} \frac{d\hat\chi_{\bf k}}{dN} e^{i \b k \cdot \b x} \theta \left( k_\sigma(\tau) - k \right).
\end{align}
These coarse-grained fields are related by the equation of motion
\begin{align}
\frac{d\hat\chi_l}{dN} &= \hat v_l + \hat\xi_\chi, \\
\frac{d\hat v_l}{dN} &= -(3-\epsilon) \hat v_l - \frac{1}{{\cal H}^2} (m^2 a^2 + \nabla^2) \hat\chi_l + \hat\xi_v,
\end{align}
where use has been made of eq.~\eqref{chi-eom-free} in the second line, and the noise operators $\hat\xi_\chi$ and $\hat\xi_v$ that appear above are given by
\begin{align}
\hat\xi_\chi(\tau,\b x) &= [1-\epsilon(\tau)] \int \frac{d^3k}{(2\pi)^{3/2}} \hat\chi_{\bf k}(\tau) e^{i \b k \cdot \b x} \delta \left( \frac{k}{k_\sigma(\tau)} - 1 \right), \\
\hat\xi_v(\tau,\b x) &= [1-\epsilon(\tau)] \int \frac{d^3k}{(2\pi)^{3/2}} \frac{d\hat\chi_{\bf k}}{dN} e^{i \b k \cdot \b x} \delta \left( \frac{k}{k_\sigma(\tau)} - 1 \right).
\end{align}

In the following we will be interested in a single patch and ignore the spatial variation of the fields, hence dropping the $\b x$-dependence.  Then only the time label of the fields remains, for which we switch to the $e$-folding time $N=\log a$.  The commutators of the noise operators are then given by
\begin{equation}\label{[,]-sharp}
[\hat\xi_\chi(N_1), \hat\xi_\chi(N_2)] = 0, \qquad [\hat\xi_v(N_1), \hat\xi_v(N_2)] = 0,
\end{equation}
\begin{equation}\label{[chi,v]-sharp}
[\hat\xi_\chi(N_1), \hat\xi_v(N_2)] = 2i \sigma^3 (1-\epsilon)\left( \frac{H}{2\pi} \right)^2 \delta(N_1-N_2),
\end{equation}
where $\epsilon$, $H$ and $k_\sigma$ are evaluated at $N=N_1$.  The anti-commutators, evaluated in the vacuum state, are also found to be:
\begin{equation}\label{[chi,chi]+0sharp}
\langle 0| \{ \hat\xi_\chi(N_1), \hat\xi_\chi(N_2) \} |0 \rangle = 2(1-\epsilon) {\cal P}_\chi \delta(N_1-N_2),
\end{equation}
\begin{equation}\label{[chi,v]+0sharp}
\langle 0| \{ \hat\xi_\chi(N_1), \hat\xi_v(N_2) \} |0 \rangle = 2(1-\epsilon) {\cal P}_{\chi,v} \delta(N_1-N_2),
\end{equation}
\begin{equation}\label{[v,v]+0sharp}
\langle 0| \{ \hat\xi_v(N_1), \hat\xi_v(N_2) \} |0 \rangle = 2(1-\epsilon) {\cal P}_v \delta(N_1-N_2),
\end{equation}
where 
\begin{equation}
{\cal P}_f(k,N) = \frac{k^3}{2\pi^2} |f_k(N)|^2
\end{equation}
is the dimensionless power spectrum of $f=\chi$ or $f=v=d\chi/dN$, and 
\begin{equation}
{\cal P}_{\chi,v}(k,N) = \frac{k^3}{2\pi^2} \Re \left[ \chi_k(N) v^*_k(N) \right],
\end{equation}
all of which are evaluated at $k=k_\sigma(N_1)$ and $N=N_1$.

A necessary condition for having a classical picture is that the correlation functions be real and also insensitive to the order of observables, so we demand that the commutators be much smaller than the anti-commutators.  Since the two commutators in eq.~\eqref{[,]-sharp} already vanish, we only need to ensure that the classicality factor
\begin{equation}
C = \left| \frac{\langle 0| \{ \hat\xi_\chi(N_1), \hat\xi_v(N_2) \} |0 \rangle}{\langle 0| [ \hat\xi_\chi(N_1), \hat\xi_v(N_2) ] |0 \rangle} \right|
\end{equation}
be much larger than unity.\footnote{The smallness must be in absolute value, since the anti-commutator is real and the commutator is imaginary.}  Using eqs.~\eqref{[chi,v]-sharp} and \eqref{[chi,v]+0sharp}, $C\gg1$ reads
\begin{equation}\label{classicality}
{\cal P}_{\chi,v}(k_\sigma(N),N) \gg \sigma^3 \left( \frac{H(N)}{2\pi} \right)^2.
\end{equation}
This statement is equivalent to asserting that the absolute value of the symmetric combination $\chi_k \chi'^*_k + \chi_k^* \chi'_k$ is much larger than that of the antisymmetric combination $\chi_k \chi'^*_k - \chi_k^* \chi'_k$ ($ = i/a^2$, by the Wronskian identity).\footnote{Also, in terms of the Mukhanov-Sasaki variable: $\left| \Re[u_k u_k'^*] - {\cal H} |u_k|^2 \right| \gg | \Im[u_k u_k'^*] | = \frac12$.}  In other words, $\chi_k \chi'^*_k$ is approximately real: We know it has a fixed nonzero imaginary part, but it has a much larger real part.  Under these conditions, we drop the hat sign on the field and noise operators ($\hat\chi_l$, $\hat v_l$ and $\hat \xi_{\chi,v}$) and work with the stochastic fields and noises ($\chi_l$, $v_l$ and $\xi_{\chi,v}$), which are now commuting $c$-numbers.

Another way of testing for classicality is to look at the covariance matrix.  Let us elaborate. Suppose $x_i$ ($i=1,\ldots,n$) are a set of real classical random variables, conveniently chosen to have zero mean: $\langle x_i \rangle=0$ (remember than we have $\langle \hat\xi \rangle=0$ for our noises too).  Then the $n\times n$ covariance matrix $\langle x_i x_j \rangle$ must be positive semi-definite.\footnote{Proof: Let $A_{ij} = \langle x_i x_j \rangle$ and $a$ be any real-valued vector.  Then $a^T A a = \sum_{i,j=1}^n \langle a_i x_i a_j x_j \rangle = ( \sum_{i=1}^n \langle a_i x_i \rangle )^2 \geq 0$.}  Indeed, given any positive semi-definite matrix, there exist $n$ Gaussian random variables whose covariance matrix is equal to the given matrix.  Now we are given the matrix of vacuum expectation values
\begin{equation}
\begin{pmatrix}
\langle 0| \hat\xi_\chi(N_1) \hat\xi_\chi(N_2) |0 \rangle & \langle 0| \hat\xi_\chi(N_1) \hat\xi_v(N_2) |0 \rangle \\
\langle 0| \hat\xi_v(N_1) \hat\xi_\chi(N_2) |0 \rangle & \langle 0| \hat\xi_v(N_1) \hat\xi_v(N_2) |0 \rangle
\end{pmatrix},
\end{equation}
and we wish to convert it to and interpret it as the covariance matrix
\begin{equation}
\begin{pmatrix}
\langle \xi_\chi(N_1) \xi_\chi(N_2) \rangle & \langle \xi_\chi(N_1) \xi_v(N_2) \rangle \\
\langle \xi_v(N_1) \xi_\chi(N_2) \rangle & \langle \xi_v(N_1) \xi_v(N_2) \rangle
\end{pmatrix}.
\end{equation}
The original matrix is Hermitian, but neither real nor necessarily positive semi-definite.  However, it is straightforward to check that taking the real part of the off-diagonal elements makes the matrix not only real and symmetric, but also positive semi-definite.  Thus we see that if this process (taking the real part of the off-diagonal elements) introduces little change, i.e., if the off-diagonal elements are approximately real, then we have a fairly accurate classical description.  This is clearly equivalent to the classicality criterion~\eqref{classicality}.

It is useful to mention a few remarks about the eigenvalues of the covariance matrix.  The entries are given by eqs.~\eqref{[chi,chi]+0sharp}--\eqref{[v,v]+0sharp}, except that a division by two is necessary to go from anti-commutators to correlators.  Setting aside the common factor $(1-\epsilon) \frac{k_\sigma^3}{2\pi^2} \delta(N_1-N_2)$, this matrix is of the form
\begin{equation}
\begin{pmatrix}
|\chi|^2 & \Re(\chi v^*) \\
\Re(\chi v^*) & |v|^2
\end{pmatrix}.
\end{equation}
As we mentioned above, this is a positive semi-definite matrix.  Its eigenvalues are
\begin{equation}
\lambda_\pm = \frac{T \pm \sqrt{T^2-4D}}{2},
\end{equation}
where $T = |\chi|^2 + |v|^2$ is the trace and $D = [ \Im(\chi v^*) ]^2$ ($= 1/(2{\cal H}a^2)^2$, by the Wronskian identity) is the determinant.  As mentioned above, classicality implies that $\Im(\chi v^*) \ll \Re(\chi v^*)$.  On the other hand, $|\Re(\chi v^*)| \leq |\chi| |v| \leq \frac12 (|\chi|^2 + |v|^2)$.  Thus $D \ll T^2/4$, which means that one of the eigenvalues is much smaller than the other:
\begin{equation}\label{eig-approx}
\lambda_+ \approx T \gg \lambda_- \approx \frac{D}{T}.
\end{equation}
In fact, the ratio $\lambda_+/\lambda_-$ is controlled by the square of the classicality factor, $C^2$.  This is an important result, because it means that whenever the system attains classicality, there is practically only one independent noise and the other one has a comparably tiny amplitude.

We have gone through this rather detailed review to emphasize that everything we have said so far applies to any inflationary background, i.e., any scale factor $a(\tau)$ with $\ddot a>0$, and not just the dS or quasi-dS background.  We will see below how particular choices of $w$ lead to different consequences.  Finally, let us write down the well-known Langevin equations for the long-mode field:
\begin{align}
\frac{d\chi_l}{dN} &= v_l + \xi_\chi, \label{Langevin-chi}\\
\frac{dv_l}{dN} &= -(3-\epsilon) v_l - \frac{m^2}{H^2} \chi_l + \xi_v, \label{Langevin-v}
\end{align}
where we have dropped the gradient term as we work in a single patch at fixed $\b x$.  To study this system of stochastic differential equations, we need to know the statistical properties of the noise.  In the forthcoming sections we investigate classicality in the dS case $w=-1$ and then the more general case of $w$.

\section{Free Field on dS}\label{sec:dS}

We first revisit the case of a free field on an exact de Sitter background, before considering the situation away from quasi-de~Sitter space in the next section.

In dS space with constant Hubble parameter $H$, we have $a = -1/H\tau$ and the Mukhanov-Sasaki equation reads
\begin{equation}\label{Mukhanov-Sasaki-dS}
u''_k + \left( k^2-\frac{\nu^{2}-\frac14}{\tau^2} \right) u_k = 0,
\end{equation}
where
\begin{equation}\label{nu-dS}
\nu = \sqrt{\frac{9}{4}-\frac{m^2}{H^2}}
\end{equation}
is a constant parameter.  The solution of this equation after imposing the Bunch-Davies initial condition is
\begin{equation}\label{dS-u}
u_k = \frac{1}{2} e^{i(2\nu+1)\pi/4} \sqrt{-\pi\tau}H_{\nu}(-k\tau),
\end{equation}
where $H_{\nu}$ is the Hankel function of the first kind and order $\nu$, and the irrelevant overall phase can be discarded.

\begin{figure}
\centering
\includegraphics[scale=.8]{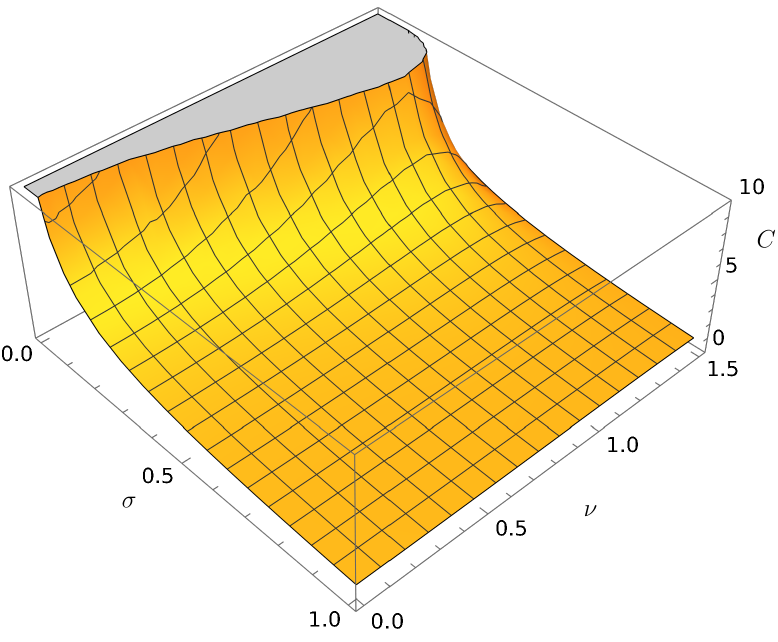}
\caption{The classicality factor $C$, the LHS/RHS of the inequality~\eqref{cl-dS-exact}, as a function of $\sigma$ and $\nu$.  The clipped region goes up all over to infinity.}
\label{fig:cl-dS}
\end{figure}

Inserting the mode function~\eqref{dS-u} in eqs.~\eqref{[chi,chi]+0sharp}--\eqref{[v,v]+0sharp} with $\epsilon=0$ and dividing by 2, we find the following stochastic correlators:
\begin{equation}\label{dS<chichi>exact}
\langle \xi_\chi(N_1) \xi_\chi(N_2) \rangle = \frac{\sigma^3 H^2}{8\pi} \left| H_{\nu}(\sigma) \right|^2 \delta(N_1-N_2),
\end{equation}
\begin{equation}\label{dS<chiv>exact}
\langle \xi_\chi(N_1) \xi_v(N_2) \rangle = -\frac{\sigma^3 H^2}{8 \pi} \left[ \left( \frac32 - \nu \right) \left| H_{\nu}(\sigma) \right|^2 + \sigma \Re[ H_{\nu}(\sigma) H_{\nu-1}^*(\sigma) ] \right] \delta(N_1-N_2),
\end{equation}
\begin{equation}\label{dS<vv>exact}
\langle \xi_v(N_1) \xi_v(N_2) \rangle = \frac{\sigma^3 H^2}{8 \pi} \left| \left( \frac32-\nu \right) H_{\nu}(\sigma) + \sigma H_{\nu-1}(\sigma) \right|^2 \delta(N_1-N_2).
\end{equation}
Thus the classicality criterion~\eqref{classicality} becomes
\begin{equation}\label{cl-dS-exact}
\left( \frac32 - \nu \right) \left| H_{\nu}(\sigma) \right|^2 + \sigma \Re[ H_{\nu}(\sigma) H_{\nu-1}^*(\sigma) ] \gg \frac{2}{\pi}.
\end{equation}
The classicality factor $C$ is equal to the ratio of the two sides (LHS/RHS) of this inequality.  It is plotted in fig.~\ref{fig:cl-dS} and it is clear that $C\gg1$ requires $\sigma\to0$.  This confirms the common claim that the wavelength cutoff of stochastic inflation has to be much larger than the horizon size for the entire range $\frac32>\nu>0$.

In the $\sigma\to0$ limit, and provided that $\nu$ is not too close to $0$ nor $\frac32$, the coefficients of the delta function on the right hand sides of eqs.~\eqref{dS<chichi>exact}--\eqref{dS<vv>exact} become, respectively,\footnote{The subleading terms are of order $\sigma^{5-2\nu}$, $\sigma^3$ and $\sigma^{3+2\nu}$, which are all negligible as long as $\nu$ is not too close to zero.}
\begin{equation}\label{dS<chichi>nu}
\left( \frac{\sigma}{2} \right)^{3-2\nu} \frac{[\Gamma(\nu) H]^2}{\pi^3},
\end{equation}
\begin{equation}\label{dS<chiv>nu}
- \left( \frac{\sigma}{2} \right)^{3-2\nu} \frac{(\frac32 - \nu) [\Gamma(\nu) H]^2}{\pi^3},
\end{equation}
\begin{equation}\label{dS<vv>nu}
\left( \frac{\sigma}{2} \right)^{3-2\nu} \frac{[(\frac32 - \nu) \Gamma(\nu) H]^2}{\pi^3}.
\end{equation}
Assuming the hierarchy mentioned in the Introduction, namely,
\begin{equation}\label{sigma-interval}
\exp \left(-\frac{3H^2}{m^2} \right) \ll \sigma^2 \ll \frac{m^2}{3H^2} \ll 1,
\end{equation}
the leading terms in the correlators become
\begin{equation}\label{dS<chichi>m}
\langle \xi_\chi(N_1) \xi_\chi(N_2) \rangle = \left( \frac{H}{2\pi} \right)^2 \delta(N_1-N_2),
\end{equation}
\begin{equation}\label{dS<chiv>m}
\langle \xi_\chi(N_1) \xi_v(N_2) \rangle = -\frac{m^2}{3H^2} \left( \frac{H}{2\pi} \right)^2 \delta(N_1-N_2),
\end{equation}
\begin{equation}\label{dS<vv>m}
\langle \xi_v(N_1) \xi_v(N_2) \rangle = \left( \frac{m^2}{3H^2} \right)^2 \left( \frac{H}{2\pi} \right)^2 \delta(N_1-N_2).
\end{equation}
As a double check, we notice that the commutator is negligible compared to the correlator in eq.~\eqref{dS<chiv>m}, if $m^2/3H^2 \gg \sigma^3$, which is a consequence of the assumed hierarchy~\eqref{sigma-interval} and consistent with  ref.~\cite{Grain:2017dqa}.  But note that the leftmost inequality, $\exp(-3H^2/m^2) \ll \sigma^2$ is not a requirement of classicality.  It is there to guarantee the additional nice property that the noise amplitudes are independent of $\sigma$.  These are all consistent with the well-known results in the literature~\cite{Sasaki:1987gy}, although it seems that classicality has always been shown in the small mass limit ($\nu\approx\frac32$), where the $\sigma$-independence property shows up, too.

If we give up the hierarchy~\eqref{sigma-interval} and the desire to have $\sigma$-independent amplitude, then the sheer requirement of classicality yields (still for $\nu$ not too close to $0$ nor $\frac32$, so using eqs.~\eqref{dS<chichi>nu}--\eqref{dS<vv>nu}):
\begin{equation}
\left( \frac{\sigma}{2} \right)^{2\nu} \ll \frac1{2\pi} \left( \frac32 - \nu \right) \Gamma(\nu)^2.
\end{equation}
This is always satisfiable by a suitable choice of $\sigma$.  Indeed, since $\nu$ is away from $0$ and $\frac32$, the right hand side is of order one, thus the classicality criterion in this range effectively becomes $\sigma\ll1$, without necessarily requiring an exponential lower bound like~\eqref{sigma-interval} on $\sigma$.

Calculation of the higher order correlators beyond two-point for this non-interacting field proceeds by application of the Wick theorem and one finds that $\xi_\chi$ and $\xi_v$ are Gaussian white noises.  Furthermore, by diagonalizing the covariance matrix, it is evident that there is only one non-zero eigenvalue, corresponding to one independent Gaussian white noise with normalized unit amplitude, which we denote by $\xi_n$, that satisfies
\begin{equation}
\langle \xi_n(N_1) \xi_n(N_2) \rangle = \delta(N_1-N_2).
\end{equation}
The field noises are then given by
\begin{equation}\label{xi-vs-xi_n}
\xi_\chi = \frac{H}{2\pi} \xi_n, \qquad \xi_v = - \frac{m^2}{3H^2} \frac{H}{2\pi} \xi_n.
\end{equation}

Now let us consider the case where $\nu$ is very close to $\frac32$, i.e., the small mass limit.  Clearly, the factor $\frac32-\nu$ in eqs.~\eqref{dS<chiv>nu} and \eqref{dS<vv>nu} vanishes and higher order terms in $\sigma$ must be included.  Also, for the massless case ($m=0$), it is impossible to choose $\sigma$ to satisfy the inequalities in~\eqref{sigma-interval}.  Nevertheless, when $\nu=\frac32$, we have the exact result (to all orders in $\sigma$) for eqs.~\eqref{dS<chichi>exact}--\eqref{dS<vv>exact}:
\begin{equation}\label{dS<chichi>m=0}
\langle \xi_\chi(N_1) \xi_\chi(N_2) \rangle = (1+\sigma^2) \left( \frac{H}{2\pi} \right)^2 \delta(N_1-N_2),
\end{equation}
\begin{equation}\label{dS<chiv>m=0}
\langle \xi_\chi(N_1) \xi_v(N_2) \rangle = -\sigma^2 \left( \frac{H}{2\pi} \right)^2 \delta(N_1-N_2),
\end{equation}
\begin{equation}\label{dS<vv>m=0}
\langle \xi_v(N_1) \xi_v(N_2) \rangle = \sigma^4 \left( \frac{H}{2\pi} \right)^2 \delta(N_1-N_2).
\end{equation}
Clearly, this is still a classical stochastic situation if $\sigma\ll1$, as the commutators are much smaller than the anti-commutators, although the $\xi_v$ noise amplitude depends on $\sigma$.  Furthermore, the absence of interactions still implies Gaussianity.  However, there is no longer an exactly vanishing eigenvalue, but to the leading order we can write
\begin{equation}\label{xi-vs-xi_n-m0}
\xi_\chi = \frac{H}{2\pi} \xi_n, \qquad \xi_v = -\sigma^2 \frac{H}{2\pi} \xi_n,
\end{equation}
as the effect of a second independent noise starts at the order $\sigma^3$, which is why it is practically irrelevant.  Thus, if $m=0$, we have essentially no noise on $v_l$ in the $\sigma\to0$ limit, i.e., for every realization of the stochastic field $\chi_l$, the field $v_l$ is deterministically specified by eq.~\eqref{Langevin-v} without the $\xi_v$ term.

Finally, at $\nu=0$, corresponding to $m=3H/2$, the $\sigma\to0$ limit of the correlators~\eqref{dS<chichi>exact}--\eqref{dS<vv>exact} becomes:
\begin{equation}
\langle \xi_\chi(N_1) \xi_\chi(N_2) \rangle = \frac{\sigma^3 (\log\sigma)^2 H^2}{2\pi^3} \delta(N_1-N_2),
\end{equation}
\begin{equation}\label{dS<chiv>nu=0}
\langle \xi_\chi(N_1) \xi_v(N_2) \rangle = -\frac32 \frac{\sigma^3 (\log\sigma)^2 H^2}{2\pi^3} \delta(N_1-N_2),
\end{equation}
\begin{equation}
\langle \xi_v(N_1) \xi_v(N_2) \rangle = \frac94 \frac{\sigma^3 (\log\sigma)^2 H^2}{2\pi^3} \delta(N_1-N_2).
\end{equation}
Thus the measure of classicality (smallness of the commutator compared to eq.~\eqref{dS<chiv>nu=0}) is equivalent to the largeness of $(\log\sigma)^2$.  This means that $\sigma$ must be exponentially small.  For example, to achieve a level of classicality that is a thousand times larger than the quantumness ($C=1000$), we need $\sigma_{1000} \approx \exp(-\sqrt{1000\pi/3}) \sim 10^{-14}$.  Furthermore, the noises are Gaussian with only one nonzero eigenvalue, so they are both proportional to a single normalized Gaussian noise $\xi_n$:
\begin{equation}
\xi_\chi = \left( \frac{\sigma}{2\pi} \right)^{3/2} \left| \log(\sigma^2) \right| H \xi_n, \qquad \xi_v = -\frac32 \left( \frac{\sigma}{2\pi} \right)^{3/2} \left| \log(\sigma^2) \right| H \xi_n.
\end{equation}
These amplitudes are $\sigma$-dependent but minuscule.  So there is practically no noise and we have a deterministic classical system.  This observation confirms the prior expectation that heavier fields behave more classical than lighter ones.

\section{Free Massless Field on Accelerating FLRW}\label{sec:w}

\begin{figure}
\centering
\includegraphics[scale=.8]{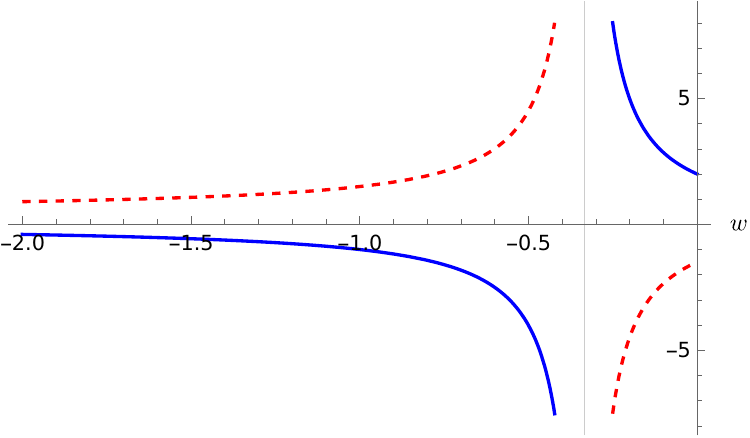}
\caption{Plots of $p$ (solid blue) and $\nu$ (dashed red) as functions of $w$ from eqs.~\eqref{p} and \eqref{nu-w}.  The relevant range is $w<-\frac13$ (to the left of the grey vertical line, where both quantities diverge).  Notable values of the triplet $(w,p,\nu)$ are $(-1,-1,\frac32)$ and $(-\frac53, -\frac12, 1)$.  As $w\to-\infty$ (which is, by the way, highly nonphysical), we have $p\to0$ and $\nu\to\frac12$.}
\label{fig:pnw}
\end{figure}

In the previous section we worked in an exact dS background.  Now we consider a massless field on an FLRW background whose matter content is a perfect fluid  whose equation of motion parameter $w$ is time-independent.    Of course, as we emphasized before, we also require an inflationary FLRW spacetime, so $w<-\frac13$.  The evolution of the scale factor can be easily found by solving Friedmann equations and we obtain:
\begin{equation}\label{p}
\frac{a(\tau)}{a_0} = \left( \frac{\tau}{\tau_0} \right)^p, \qquad \text{where} \qquad p = \frac{2}{1+3w},
\end{equation}
\begin{equation}
\epsilon = 1 + \frac1p = \frac32 (1+w),
\end{equation}
\begin{equation}
H(\tau) = \frac{p}{\tau a(\tau)} = H_0 \left( \frac{a(\tau)}{a_0} \right)^{-\frac32 (1+w)} = H_0 e^{-\epsilon N},
\end{equation}
where $\tau_0$ is an arbitrary reference time used as the origin $N=0$ of the $e$-folds and at which the scale factor and the Hubble are set by $a_0$ and $H_0$, respectively.  Fig.~\ref{fig:pnw} includes a plot of $p$ as a functions of $w$.

As before, we have the formulas for the noise at our disposal from section~\ref{sec:rev}.  The only non-vanishing commutator can be obtained from eq.~\eqref{[chi,v]-sharp} which now yields
\begin{equation}\label{comm-w}
[\hat\xi_\chi(N_1), \hat\xi_v(N_2)] = -i\sigma^3(1+3w) \left( \frac{H}{2\pi} \right)^2 \delta(N_1-N_2),
\end{equation}
where $H=H_0 \exp(-\epsilon N_1)$ is the time-dependent Hubble parameter.

The correlators are to be read from the anti-commutators given in eqs.~\eqref{[chi,chi]+0sharp}--\eqref{[v,v]+0sharp}.  To obtain the mode function, we notice that the Mukhanov-Sasaki equation is again of the form~\eqref{Mukhanov-Sasaki-dS}, thanks to the fact that $m=0$ and the observation that
\begin{equation}
\frac{a''}{a} = \frac{\nu^2-\frac14}{\tau^2},
\end{equation}
where\footnote{Actually, the absolute values are unnecessary in the range of interest $w<-1/3$.}
\begin{equation}\label{nu-w}
\nu = \left| \frac12 - p \right| = \frac32 \left| \frac{w-1}{1+3w} \right|.
\end{equation}
Note that $m=0$ is crucial in deriving $a''/a\propto1/\tau^2$, and the presence of mass would ruin this property.  Also, as a side remark, note that a qualitative difference with the previous section is that $\nu\geq\frac32$ here, whereas $\nu\leq\frac32$ there.  For reference, we have included a plot of $\nu$ as a function of $w$ in fig.~\ref{fig:pnw}.  Since $\nu$ is a constant independent of $\tau$, we have the same solution~\eqref{dS-u} as in the previous section, except that $\nu$ is now given by eq.~\eqref{nu-w} instead of eq.~\eqref{nu-dS}.  Bearing in mind that $k_\sigma = \sigma {\cal H} = p\sigma/\tau$, we find
\begin{equation}\label{<chichi>w}
\langle \xi_\chi(N_1) \xi_\chi(N_2) \rangle = \frac{\sigma^3 H^2}{8\pi} \left| H_{\nu}(-p\sigma) \right|^2 \delta(N_1-N_2),
\end{equation}
\begin{equation}\label{<chiv>w}
\langle \xi_\chi(N_1) \xi_v(N_2) \rangle = -\frac{\sigma^4 H^2}{8 \pi} \Re \left[ H_{\nu}(-p\sigma) H_{\nu-1}^*(-p\sigma) \right] \delta(N_1-N_2),
\end{equation}
\begin{equation}\label{<vv>w}
\langle \xi_v(N_1) \xi_v(N_2) \rangle = \frac{\sigma^5 H^2}{8 \pi} \left| H_{\nu-1}(-p\sigma) \right|^2 \delta(N_1-N_2).
\end{equation}
These are exact expressions for the noise amplitude that we are going to exploit in the sequel.  Before moving on, let us mention a remarkable cancellation here.  Instead of the combination
\begin{equation}
\left( \frac32-\nu \right) H_{\nu}(\sigma) + \sigma H_{\nu-1}(\sigma)
\end{equation}
that appears in eqs.~\eqref{dS<chiv>exact} and \eqref{dS<vv>exact}, we obtain
\begin{equation}
\left( 1 + \frac{2\nu-1}{2p} \right) H_{\nu}(-p\sigma) + \sigma H_{\nu-1}(-p\sigma)
\end{equation}
in eqs.~\eqref{<chiv>w} and \eqref{<vv>w}.  Since $\nu = \frac12 - p$, the term involving $H_{\nu}$ disappears altogether.  Also note that the factor $1-\epsilon$ in the noise amplitudes above is canceled out by other factors coming from the mode functions.  As a cross check, we notice that setting $w=-1$ in eqs.~\eqref{<chichi>w}--\eqref{<vv>w}, reduces them to eqs.~\eqref{dS<chichi>m=0}--\eqref{dS<vv>m=0}, which is reassuring since the latter is the case of a massless field on exact dS background.

\begin{figure}
\centering
\includegraphics[scale=.8]{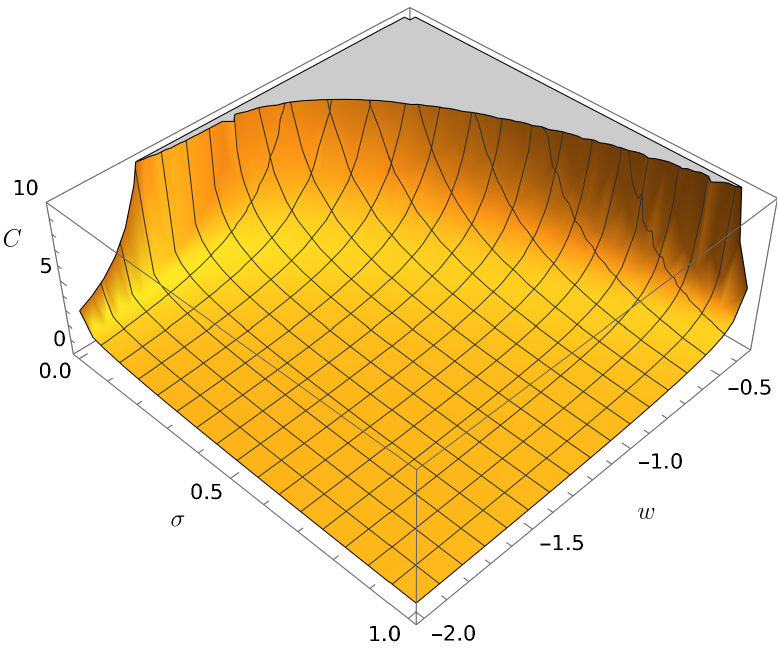}
\caption{The classicality factor $C$, the LHS/RHS of the inequality~\eqref{cl-w-exact}, as a function of $\sigma$ and $\nu$.  The clipped region goes up all over to infinity.  The upper range of $w$ is chosen to be slightly less than $-\frac13$ in this figure, since there is a blow-up right at $w=-\frac13$.}
\label{fig:cl-w}
\end{figure}

With the aid of eq.~\eqref{<chiv>w}, we now check the classicality criterion~\eqref{classicality}, which now reads
\begin{equation}\label{cl-w-exact}
-p \sigma \Re \left[ H_{\nu}(-p\sigma) H_{\nu-1}^*(-p\sigma) \right] \gg \frac{2}{\pi}.
\end{equation}
Again the classicality factor $C$ is the LHS/RHS of this inequality and is plotted in fig.~\ref{fig:cl-w}.  In contrast to the case of dS space (fig.~\ref{fig:cl-dS}), we see that it is now not necessary to have small $\sigma$ in order to achieve classicality.  In particular, $C$ diverges in the vicinity of $w=-\frac13$. 

\begin{figure}
\centering
\includegraphics[scale=.8]{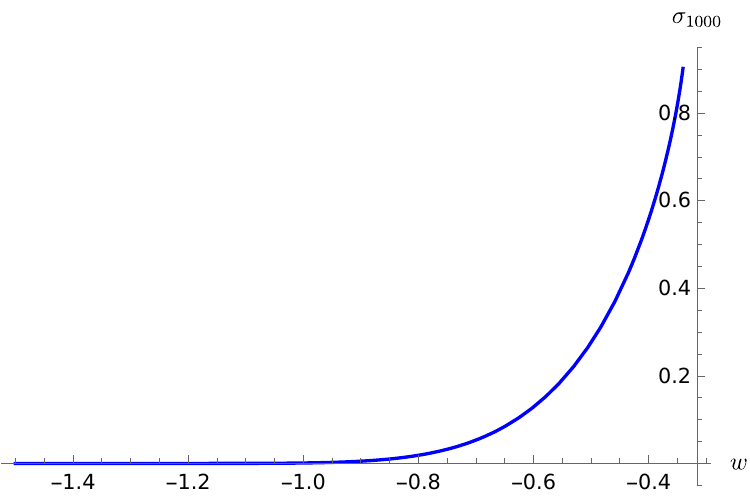}
\caption{The value of the cutoff parameter $\sigma_{1000}$ at which the classicality factor $C$ is equal to 1,000.}
\label{fig:sigma1000}
\end{figure}

To have a more quantitative picture, consider the small $\sigma$ expansion of $C$.  There are three series of terms, starting with $\sigma^{2-2\nu}$, $\sigma^0$ and $\sigma^{2\nu}$, respectively:
\begin{equation}\label{cl-w-approx}
\begin{aligned}
C &= \frac{1}{\pi} \Gamma(\nu) \Gamma(\nu-1) \left[ (\nu-\frac12) \frac{\sigma}{2} \right]^{2-2\nu} \left( 1 + O(\sigma^2) \right) \\
&- \cot(\nu\pi) \left( 1 + O(\sigma^2) \right) \\
&+ \frac{\pi}{\Gamma(\nu) \Gamma(\nu+1) \sin^2(\nu\pi)} \left[ (\nu-\frac12) \frac{\sigma}{2} \right]^{2\nu} \left( 1 + O(\sigma^2) \right).
\end{aligned}
\end{equation}
For $\nu>1$ the first term dominates and we have always the opportunity to achieve $C\gg1$, which is desired for classicality.  It is now clear that near $w=-\frac13$, where $\nu$ blows up (see fig.~\ref{fig:pnw}), it is quite easy to obtain fairly large values of $\sigma$ (of course, not larger than 1) that make $C$ large and hence the classicality criterion~\eqref{cl-w-exact} satisfied.  This is both due to the large exponent ($2-2\nu$) of $\sigma$ in eq.~\eqref{cl-w-approx}, as well as the presence of huge factorials in its coefficient.  Let us define $\sigma_{1000}$ as the cutoff value at which we can achieve a classicality factor of a thousand ($C=1000$).  Then at $w=-0.4$ we have $\sigma_{1000}=0.55$, which is equivalent to a wavelength cutoff $1/\sigma H$ equal to twice the horizon size.  Other values of $\sigma_{1000}$ are depicted in fig.~\ref{fig:sigma1000}.  Note that for $w\gtrsim-0.5$, we can achieve high levels of classicality by choosing $\sigma\sim1$.  This supports our claim that classicality does not generically require $\sigma\ll1$.

As $\nu$ decreases and falls below $1$ (corresponding to $w$ falling below $-\frac53$) the term in the second line of eq.~\eqref{cl-w-approx} dominates.  Since this term is constant in $\sigma$, it is now impossible to control the magnitude of $C$ by tuning to small values of $\sigma$.  This is also evident in fig.~\ref{fig:cl-w} where the blow-up at $\sigma=0$ disappears for $w<-\frac53$.  Of course, $-\cot(\nu\pi)$ is still large around $\nu=1$, so classicality is not lost suddenly.

The next natural question is whether we can still have $\sigma$-independent noise amplitudes.  In general, the noise amplitudes in eqs.~\eqref{<chichi>w}--\eqref{<vv>w} are $\sigma$-dependent, even when the classicality criterion is met.  There is a special limit in which $\sigma$-independence can be achieved, which we now describe.  By looking at the $\xi_\chi$ noise in eq.~\eqref{<chichi>w} we find that we need $H_\nu(-p\sigma)$ to scale like $\sigma^{-3/2}$.  Using the small argument expansion of Hankel function $H_\nu(-p\sigma) \propto \sigma^{-\nu}$, we find that this is possible around $\nu=\frac32$, i.e., in the quasi-dS regime.  Just like the conventional case, we also need to choose $\sigma$ such that $\sigma^{3-2\nu}$ is close to unity, i.e.,
\begin{equation}\label{sigma-nu}
\sigma \gg \exp \left( -\frac{1}{|2\nu-3|} \right).
\end{equation}
This is the same as the classicality condition of ref.~\cite{Grain:2017dqa}.  Under these circumstances the correlators~\eqref{<chichi>w}--\eqref{<vv>w} become
\begin{equation}\label{<chichi>w3/2}
\langle \xi_\chi(N_1) \xi_\chi(N_2) \rangle = \left[ \Gamma(\nu) \left( \frac{-2}{p} \right)^\nu \right]^2 \frac{H^2}{8\pi^3} \delta(N_1-N_2),
\end{equation}
\begin{equation}\label{<chiv>w3/2}
\langle \xi_\chi(N_1) \xi_v(N_2) \rangle = -\sigma^2  \Gamma(\nu) \Gamma(\nu-1) \left( \frac{-2}{p} \right)^{2\nu-1} \frac{H^2}{8\pi^3} \delta(N_1-N_2),
\end{equation}
\begin{equation}\label{<vv>w3/2}
\langle \xi_v(N_1) \xi_v(N_2) \rangle = \sigma^4 \left[ \Gamma(\nu-1) \left( \frac{-2}{p} \right)^{\nu-1} \right]^2 \frac{H^2}{8\pi^3} \delta(N_1-N_2).
\end{equation}
We observe that the coefficients of the delta functions are equal to those in eqs.~\eqref{dS<chichi>m=0}--\eqref{dS<vv>m=0} (the massless field on exact dS) with corrections of order $w+1$, as they should.  Comparing eq.~\eqref{<chiv>w3/2} with eq.~\eqref{comm-w}, the classicality criterion is found to be $\sigma \ll -p \approx 1$.  Thus we recover the same results as before in this special case, which is no surprise, as this is essentially the dS limit.  

Finally, we express our noises in terms of normalized Gaussian white noises.  In general, the eigenvalues of the covariance matrix formed by eqs.~\eqref{<chichi>w}--\eqref{<vv>w} are given by eq.~\eqref{eig-approx}.  In the $\sigma$-independent case of the last paragraph, the answer reduces to:
\begin{equation}
\xi_\chi = \frac{\Gamma(\nu)}{\sqrt{2\pi}} \left( \frac{-2}{p} \right)^\nu \frac{H}{2\pi} \xi_n, \qquad
\xi_\chi = -\sigma^2 \frac{\Gamma(\nu-1)}{\sqrt{2\pi}} \left( \frac{-2}{p} \right)^{\nu-1} \frac{H}{2\pi} \xi_n.
\end{equation}
In principle, one could plug this and other noises we found in this paper into the Langevin equations~\eqref{Langevin-chi} and \eqref{Langevin-v} and try to solve them directly or by converting to the corresponding Fokker-Planck equation.  This is not awkward as the time-dependence of the noise is straightforward ($\propto e^{-2\epsilon N}$), but it is not the stated purpose of this work.

\section{Summary and Discussion}\label{sec:summary}

We have revisited the question of classicality of the stochastic noise for a free (non-interacting) field on an accelerating cosmological background.    For the problem at hand, classicality reduces to the smallness of the commutators compared to the anti-commutators, and this criterion is contained in Eq.~\eqref{classicality}.  We also noticed that of the two classical noises, one has negligible amplitude and there is thus effectively only one independent noise.

We reviewed the commonly studied case of a massive field on an exact dS background in section~\ref{sec:dS} and confirmed that the cutoff $\sigma$ (used to separate IR and UV modes) must be small.  While all previous studies that we know show this for small mass, we did this for the entire range $3H/2 > m >0$.  We found that for most of the range, $\sigma\ll1$ is sufficient for classicality, but that we need $\sigma\ll(m/H)^{2/3}$ and $(\log\sigma)^2\gg1$, near $m=0$ and $m=3H/2$, respectively.  The noise amplitude depends on $\sigma$, except near $m=0$ and with the extra assumption $\sigma^2 \gg \exp(-3H^2/m^2)$, in harmony with the existing literature.  Although we didn't consider the quasi-dS case with time varying $w\approx-1$, similar results hold there too (see, for example, ref.~\cite{Grain:2017dqa}).

In section~\ref{sec:w} we studied the massless field on an accelerating background with general $w>-\frac13$, not necessarily close to the quasi-dS regime $w\approx-1$.  We obtained exact expressions for the noise amplitudes in Eqs.~\eqref{<chichi>w}--\eqref{<vv>w}, which we used to investigate the classicality criterion.  Our key observation was that it is not necessary to have $\sigma\ll1$ in order to achieve classicality, especially as $w=-\frac13$ is approached.  This is a novel feature compared to the standard lore near $w=-1$.  We emphasize that there is nothing intrinsic about $\sigma<1$ that connects it to classicality.  In fact, there is another approach to stochastic inflation that employs fields coarse-grained on a causally connected subhorizon region and it also yields a classical stochastic picture \cite{Mirbabayi:2020vyt}; in terms of Starobinsky's approach adopted here, this amounts to $\sigma>1$. 

We also found that as long as $w>-\frac53$, it is always possible to arrange for classicality by a suitable choice of $\sigma$.  Of course, $w<-1$ is in conflict with energy conditions and is not well-motivated theoretically.  But given some observational motivations for $w$ slightly less than $-1$, we didn't exert these extra constraints on our analysis to limit its domain.

A nice feature of the conventional analysis in the quasi-dS regime is that the noise amplitudes are independent of the cutoff $\sigma$.  We recover the cutoff-independence condition of ref.~\cite{Grain:2017dqa} near $w=-1$ in eq.~\eqref{sigma-nu}, but this feature is lost away from $w=-1$.  Although cutoff-independence is an attractive property of a physical observable, the dependence of a quantity on its scale of coarse-graining does not by itself make it irrelevant to the calculation of observables.  So the situation here is that we have a quantity, namely the coarse-grained field, whose definition depends on the cutoff scale $\sigma$ (through the window function).  The dynamics of this quantity also depends on $\sigma$ through the noise amplitude.  Except for certain limits where this dependence is absent, there is no universal $\sigma$-independent dynamics.  It would be most interesting to come up with physical observables that are sensitive to the field coarse-grained on the scale $\sigma$ for which our $\sigma$-dependent amplitude determines the dynamics.

We should also mention the assumptions under which our analysis is performed.  In the course of the derivation of the results of section~\ref{sec:rev} and then  throughout the paper, we have assumed the Bunch-Davies vacuum as the state in which our expectation values are evaluated.  We have also used a sharp window function (Heaviside step function) to define the coarse-grained fields, which is the origin of the whiteness of the noise.  In addition, our field was free which simplified the complications that would arise from interaction and led to trivial higher moments of noise and its Gaussian statistics.  Modification of any of these assumptions can change our analytical results, but the major conclusion cannot be changed as we have already found an example of classicality with $\sigma\sim1$.

\section*{Acknowledgements}

I acknowledge financial support from the research council of University of Tehran.
	
\bibliography{classicality-w}

\providecommand{\href}[2]{#2}\begingroup\raggedright\begin{thebibliography}{10}

\bibitem{Guth:1980zm}
A.H.~Guth, \emph{{The Inflationary Universe: A Possible Solution to the Horizon
  and Flatness Problems}},
  \href{https://doi.org/10.1103/PhysRevD.23.347}{\emph{Phys. Rev. D} {\bfseries
  23} (1981) 347}.

\bibitem{Linde:1981mu}
A.D.~Linde, \emph{{A New Inflationary Universe Scenario: A Possible Solution of
  the Horizon, Flatness, Homogeneity, Isotropy and Primordial Monopole
  Problems}}, \href{https://doi.org/10.1016/0370-2693(82)91219-9}{\emph{Phys.
  Lett. B} {\bfseries 108} (1982) 389}.

\bibitem{Albrecht:1982wi}
A.~Albrecht and P.J.~Steinhardt, \emph{{Cosmology for Grand Unified Theories
  with Radiatively Induced Symmetry Breaking}},
  \href{https://doi.org/10.1103/PhysRevLett.48.1220}{\emph{Phys. Rev. Lett.}
  {\bfseries 48} (1982) 1220}.

\bibitem{Vilenkin:1983xp}
A.~Vilenkin, \emph{{Quantum Fluctuations in the New Inflationary Universe}},
  \href{https://doi.org/10.1016/0550-3213(83)90208-0}{\emph{Nucl. Phys. B}
  {\bfseries 226} (1983) 527}.

\bibitem{Vilenkin:1983xq}
A.~Vilenkin, \emph{{The Birth of Inflationary Universes}},
  \href{https://doi.org/10.1103/PhysRevD.27.2848}{\emph{Phys. Rev. D}
  {\bfseries 27} (1983) 2848}.

\bibitem{Linde:1986fd}
A.D.~Linde, \emph{{Eternally Existing Selfreproducing Chaotic Inflationary
  Universe}}, \href{https://doi.org/10.1016/0370-2693(86)90611-8}{\emph{Phys.
  Lett. B} {\bfseries 175} (1986) 395}.

\bibitem{Starobinsky:1986fx}
A.A.~Starobinsky, \emph{{STOCHASTIC DE SITTER (INFLATIONARY) STAGE IN THE EARLY
  UNIVERSE}}, \href{https://doi.org/10.1007/3-540-16452-9_6}{\emph{Lect. Notes
  Phys.} {\bfseries 246} (1986) 107}.

\bibitem{Rey:1986zk}
S.-J.~Rey, \emph{{Dynamics of Inflationary Phase Transition}},
  \href{https://doi.org/10.1016/0550-3213(87)90058-7}{\emph{Nucl. Phys. B}
  {\bfseries 284} (1987) 706}.

\bibitem{Aryal:1987vn}
M.~Aryal and A.~Vilenkin, \emph{{The Fractal Dimension of Inflationary
  Universe}}, \href{https://doi.org/10.1016/0370-2693(87)90932-4}{\emph{Phys.
  Lett. B} {\bfseries 199} (1987) 351}.

\bibitem{Starobinsky:1994bd}
A.A.~Starobinsky and J.~Yokoyama, \emph{{Equilibrium state of a selfinteracting
  scalar field in the De Sitter background}},
  \href{https://doi.org/10.1103/PhysRevD.50.6357}{\emph{Phys. Rev. D}
  {\bfseries 50} (1994) 6357}
  [\href{https://arxiv.org/abs/astro-ph/9407016}{{\ttfamily
  astro-ph/9407016}}].

\bibitem{Sasaki:1987gy}
M.~Sasaki, Y.~Nambu and K.-i.~Nakao, \emph{{Classical Behavior of a Scalar
  Field in the Inflationary Universe}},
  \href{https://doi.org/10.1016/0550-3213(88)90132-0}{\emph{Nucl. Phys. B}
  {\bfseries 308} (1988) 868}.

\bibitem{Nambu:1987ef}
Y.~Nambu and M.~Sasaki, \emph{{Stochastic Stage of an Inflationary Universe
  Model}}, \href{https://doi.org/10.1016/0370-2693(88)90974-4}{\emph{Phys.
  Lett. B} {\bfseries 205} (1988) 441}.

\bibitem{Nambu:1988je}
Y.~Nambu and M.~Sasaki, \emph{{Stochastic approach to chaotic inflation and the
  distribution of universes}},
  \href{https://doi.org/10.1016/0370-2693(89)90385-7}{\emph{Phys. Lett. B}
  {\bfseries 219} (1989) 240}.

\bibitem{Kandrup:1988sc}
H.E.~Kandrup, \emph{{STOCHASTIC INFLATION AS A TIME DEPENDENT RANDOM WALK}},
  \href{https://doi.org/10.1103/PhysRevD.39.2245}{\emph{Phys. Rev. D}
  {\bfseries 39} (1989) 2245}.

\bibitem{Nakao:1988yi}
K.-i.~Nakao, Y.~Nambu and M.~Sasaki, \emph{{Stochastic Dynamics of New
  Inflation}}, \href{https://doi.org/10.1143/PTP.80.1041}{\emph{Prog. Theor.
  Phys.} {\bfseries 80} (1988) 1041}.

\bibitem{Nambu:1989uf}
Y.~Nambu, \emph{{Stochastic Dynamics of an Inflationary Model and Initial
  Distribution of Universes}},
  \href{https://doi.org/10.1143/PTP.81.1037}{\emph{Prog. Theor. Phys.}
  {\bfseries 81} (1989) 1037}.

\bibitem{Mollerach:1990zf}
S.~Mollerach, S.~Matarrese, A.~Ortolan and F.~Lucchin, \emph{{Stochastic
  inflation in a simple two field model}},
  \href{https://doi.org/10.1103/PhysRevD.44.1670}{\emph{Phys. Rev. D}
  {\bfseries 44} (1991) 1670}.

\bibitem{Spokoiny:1993uc}
B.~Spokoiny, \emph{{Stochastic nonde Sitter inflation}},
  \href{https://doi.org/10.1016/0083-6656(93)90080-4}{\emph{Vistas Astron.}
  {\bfseries 37} (1993) 481}
  [\href{https://arxiv.org/abs/hep-th/9305159}{{\ttfamily hep-th/9305159}}].

\bibitem{Linde:1993xx}
A.D.~Linde, D.A.~Linde and A.~Mezhlumian, \emph{{From the Big Bang theory to
  the theory of a stationary universe}},
  \href{https://doi.org/10.1103/PhysRevD.49.1783}{\emph{Phys. Rev. D}
  {\bfseries 49} (1994) 1783}
  [\href{https://arxiv.org/abs/gr-qc/9306035}{{\ttfamily gr-qc/9306035}}].

\bibitem{Linde:1996hg}
A.D.~Linde, D.A.~Linde and A.~Mezhlumian, \emph{{Nonperturbative amplifications
  of inhomogeneities in a selfreproducing universe}},
  \href{https://doi.org/10.1103/PhysRevD.54.2504}{\emph{Phys. Rev. D}
  {\bfseries 54} (1996) 2504}
  [\href{https://arxiv.org/abs/gr-qc/9601005}{{\ttfamily gr-qc/9601005}}].

\bibitem{Kunze:2006tu}
K.E.~Kunze, \emph{{Perturbations in stochastic inflation}},
  \href{https://doi.org/10.1088/1475-7516/2006/07/014}{\emph{JCAP} {\bfseries
  07} (2006) 014} [\href{https://arxiv.org/abs/astro-ph/0603575}{{\ttfamily
  astro-ph/0603575}}].

\bibitem{Prokopec:2007ak}
T.~Prokopec, N.C.~Tsamis and R.P.~Woodard, \emph{{Stochastic Inflationary
  Scalar Electrodynamics}},
  \href{https://doi.org/10.1016/j.aop.2007.08.008}{\emph{Annals Phys.}
  {\bfseries 323} (2008) 1324}
  [\href{https://arxiv.org/abs/0707.0847}{{\ttfamily 0707.0847}}].

\bibitem{Prokopec:2008gw}
T.~Prokopec, N.C.~Tsamis and R.P.~Woodard, \emph{{Two loop stress-energy tensor
  for inflationary scalar electrodynamics}},
  \href{https://doi.org/10.1103/PhysRevD.78.043523}{\emph{Phys. Rev. D}
  {\bfseries 78} (2008) 043523}
  [\href{https://arxiv.org/abs/0802.3673}{{\ttfamily 0802.3673}}].

\bibitem{Tsamis:2005hd}
N.C.~Tsamis and R.P.~Woodard, \emph{{Stochastic quantum gravitational
  inflation}},
  \href{https://doi.org/10.1016/j.nuclphysb.2005.06.031}{\emph{Nucl. Phys. B}
  {\bfseries 724} (2005) 295}
  [\href{https://arxiv.org/abs/gr-qc/0505115}{{\ttfamily gr-qc/0505115}}].

\bibitem{Enqvist:2008kt}
K.~Enqvist, S.~Nurmi, D.~Podolsky and G.I.~Rigopoulos, \emph{{On the
  divergences of inflationary superhorizon perturbations}},
  \href{https://doi.org/10.1088/1475-7516/2008/04/025}{\emph{JCAP} {\bfseries
  04} (2008) 025} [\href{https://arxiv.org/abs/0802.0395}{{\ttfamily
  0802.0395}}].

\bibitem{Finelli:2008zg}
F.~Finelli, G.~Marozzi, A.A.~Starobinsky, G.P.~Vacca and G.~Venturi,
  \emph{{Generation of fluctuations during inflation: Comparison of stochastic
  and field-theoretic approaches}},
  \href{https://doi.org/10.1103/PhysRevD.79.044007}{\emph{Phys. Rev. D}
  {\bfseries 79} (2009) 044007}
  [\href{https://arxiv.org/abs/0808.1786}{{\ttfamily 0808.1786}}].

\bibitem{Finelli:2010sh}
F.~Finelli, G.~Marozzi, A.A.~Starobinsky, G.P.~Vacca and G.~Venturi,
  \emph{{Stochastic growth of quantum fluctuations during slow-roll
  inflation}}, \href{https://doi.org/10.1103/PhysRevD.82.064020}{\emph{Phys.
  Rev. D} {\bfseries 82} (2010) 064020}
  [\href{https://arxiv.org/abs/1003.1327}{{\ttfamily 1003.1327}}].

\bibitem{Garbrecht:2013coa}
B.~Garbrecht, G.~Rigopoulos and Y.~Zhu, \emph{{Infrared correlations in de
  Sitter space: Field theoretic versus stochastic approach}},
  \href{https://doi.org/10.1103/PhysRevD.89.063506}{\emph{Phys. Rev. D}
  {\bfseries 89} (2014) 063506}
  [\href{https://arxiv.org/abs/1310.0367}{{\ttfamily 1310.0367}}].

\bibitem{Garbrecht:2014dca}
B.~Garbrecht, F.~Gautier, G.~Rigopoulos and Y.~Zhu, \emph{{Feynman Diagrams for
  Stochastic Inflation and Quantum Field Theory in de Sitter Space}},
  \href{https://doi.org/10.1103/PhysRevD.91.063520}{\emph{Phys. Rev. D}
  {\bfseries 91} (2015) 063520}
  [\href{https://arxiv.org/abs/1412.4893}{{\ttfamily 1412.4893}}].

\bibitem{Burgess:2014eoa}
C.P.~Burgess, R.~Holman, G.~Tasinato and M.~Williams, \emph{{EFT Beyond the
  Horizon: Stochastic Inflation and How Primordial Quantum Fluctuations Go
  Classical}}, \href{https://doi.org/10.1007/JHEP03(2015)090}{\emph{JHEP}
  {\bfseries 03} (2015) 090} [\href{https://arxiv.org/abs/1408.5002}{{\ttfamily
  1408.5002}}].

\bibitem{Burgess:2015ajz}
C.P.~Burgess, R.~Holman and G.~Tasinato, \emph{{Open EFTs, IR effects
  \textbackslash{}\& late-time resummations: systematic corrections in
  stochastic inflation}},
  \href{https://doi.org/10.1007/JHEP01(2016)153}{\emph{JHEP} {\bfseries 01}
  (2016) 153} [\href{https://arxiv.org/abs/1512.00169}{{\ttfamily
  1512.00169}}].

\bibitem{Boyanovsky:2015tba}
D.~Boyanovsky, \emph{{Effective field theory during inflation: Reduced density
  matrix and its quantum master equation}},
  \href{https://doi.org/10.1103/PhysRevD.92.023527}{\emph{Phys. Rev. D}
  {\bfseries 92} (2015) 023527}
  [\href{https://arxiv.org/abs/1506.07395}{{\ttfamily 1506.07395}}].

\bibitem{Boyanovsky:2015jen}
D.~Boyanovsky, \emph{{Effective field theory during inflation. II. Stochastic
  dynamics and power spectrum suppression}},
  \href{https://doi.org/10.1103/PhysRevD.93.043501}{\emph{Phys. Rev. D}
  {\bfseries 93} (2016) 043501}
  [\href{https://arxiv.org/abs/1511.06649}{{\ttfamily 1511.06649}}].

\bibitem{Fujita:2017lfu}
T.~Fujita and I.~Obata, \emph{{Does anisotropic inflation produce a small
  statistical anisotropy?}},
  \href{https://doi.org/10.1088/1475-7516/2018/01/049}{\emph{JCAP} {\bfseries
  01} (2018) 049} [\href{https://arxiv.org/abs/1711.11539}{{\ttfamily
  1711.11539}}].

\bibitem{Glavan:2017jye}
D.~Glavan, T.~Prokopec and A.A.~Starobinsky, \emph{{Stochastic dark energy from
  inflationary quantum fluctuations}},
  \href{https://doi.org/10.1140/epjc/s10052-018-5862-5}{\emph{Eur. Phys. J. C}
  {\bfseries 78} (2018) 371}
  [\href{https://arxiv.org/abs/1710.07824}{{\ttfamily 1710.07824}}].

\bibitem{Gorbenko:2019rza}
V.~Gorbenko and L.~Senatore, \emph{{$\lambda \phi^4$ in dS}},
  \href{https://arxiv.org/abs/1911.00022}{{\ttfamily 1911.00022}}.

\bibitem{Mirbabayi:2019qtx}
M.~Mirbabayi, \emph{{Infrared dynamics of a light scalar field in de Sitter}},
  \href{https://doi.org/10.1088/1475-7516/2020/12/006}{\emph{JCAP} {\bfseries
  12} (2020) 006} [\href{https://arxiv.org/abs/1911.00564}{{\ttfamily
  1911.00564}}].

\bibitem{Mirbabayi:2020vyt}
M.~Mirbabayi, \emph{{Markovian dynamics in de Sitter}},
  \href{https://doi.org/10.1088/1475-7516/2021/09/038}{\emph{JCAP} {\bfseries
  09} (2021) 038} [\href{https://arxiv.org/abs/2010.06604}{{\ttfamily
  2010.06604}}].

\bibitem{Cohen:2021fzf}
T.~Cohen, D.~Green, A.~Premkumar and A.~Ridgway, \emph{{Stochastic Inflation at
  NNLO}}, \href{https://doi.org/10.1007/JHEP09(2021)159}{\emph{JHEP} {\bfseries
  09} (2021) 159} [\href{https://arxiv.org/abs/2106.09728}{{\ttfamily
  2106.09728}}].

\bibitem{Cruces:2021iwq}
D.~Cruces and C.~Germani, \emph{{Stochastic inflation at all order in slow-roll
  parameters: Foundations}},
  \href{https://doi.org/10.1103/PhysRevD.105.023533}{\emph{Phys. Rev. D}
  {\bfseries 105} (2022) 023533}
  [\href{https://arxiv.org/abs/2107.12735}{{\ttfamily 2107.12735}}].

\bibitem{Cruces:2022imf}
D.~Cruces, \emph{{Review on Stochastic Approach to Inflation}},
  \href{https://doi.org/10.3390/universe8060334}{\emph{Universe} {\bfseries 8}
  (2022) 334} [\href{https://arxiv.org/abs/2203.13852}{{\ttfamily
  2203.13852}}].

\bibitem{Fujita:2013cna}
T.~Fujita, M.~Kawasaki, Y.~Tada and T.~Takesako, \emph{{A new algorithm for
  calculating the curvature perturbations in stochastic inflation}},
  \href{https://doi.org/10.1088/1475-7516/2013/12/036}{\emph{JCAP} {\bfseries
  12} (2013) 036} [\href{https://arxiv.org/abs/1308.4754}{{\ttfamily
  1308.4754}}].

\bibitem{Fujita:2014tja}
T.~Fujita, M.~Kawasaki and Y.~Tada, \emph{{Non-perturbative approach for
  curvature perturbations in stochastic $\delta N$ formalism}},
  \href{https://doi.org/10.1088/1475-7516/2014/10/030}{\emph{JCAP} {\bfseries
  10} (2014) 030} [\href{https://arxiv.org/abs/1405.2187}{{\ttfamily
  1405.2187}}].

\bibitem{Vennin:2015hra}
V.~Vennin and A.A.~Starobinsky, \emph{{Correlation Functions in Stochastic
  Inflation}}, \href{https://doi.org/10.1140/epjc/s10052-015-3643-y}{\emph{Eur.
  Phys. J. C} {\bfseries 75} (2015) 413}
  [\href{https://arxiv.org/abs/1506.04732}{{\ttfamily 1506.04732}}].

\bibitem{Vennin:2016wnk}
V.~Vennin, H.~Assadullahi, H.~Firouzjahi, M.~Noorbala and D.~Wands,
  \emph{{Critical Number of Fields in Stochastic Inflation}},
  \href{https://doi.org/10.1103/PhysRevLett.118.031301}{\emph{Phys. Rev. Lett.}
  {\bfseries 118} (2017) 031301}
  [\href{https://arxiv.org/abs/1604.06017}{{\ttfamily 1604.06017}}].

\bibitem{Assadullahi:2016gkk}
H.~Assadullahi, H.~Firouzjahi, M.~Noorbala, V.~Vennin and D.~Wands,
  \emph{{Multiple Fields in Stochastic Inflation}},
  \href{https://doi.org/10.1088/1475-7516/2016/06/043}{\emph{JCAP} {\bfseries
  06} (2016) 043} [\href{https://arxiv.org/abs/1604.04502}{{\ttfamily
  1604.04502}}].

\bibitem{Grain:2017dqa}
J.~Grain and V.~Vennin, \emph{{Stochastic inflation in phase space: Is slow
  roll a stochastic attractor?}},
  \href{https://doi.org/10.1088/1475-7516/2017/05/045}{\emph{JCAP} {\bfseries
  05} (2017) 045} [\href{https://arxiv.org/abs/1703.00447}{{\ttfamily
  1703.00447}}].

\bibitem{Pattison:2017mbe}
C.~Pattison, V.~Vennin, H.~Assadullahi and D.~Wands, \emph{{Quantum diffusion
  during inflation and primordial black holes}},
  \href{https://doi.org/10.1088/1475-7516/2017/10/046}{\emph{JCAP} {\bfseries
  10} (2017) 046} [\href{https://arxiv.org/abs/1707.00537}{{\ttfamily
  1707.00537}}].

\bibitem{Pattison:2018bct}
C.~Pattison, V.~Vennin, H.~Assadullahi and D.~Wands, \emph{{The attractive
  behaviour of ultra-slow-roll inflation}},
  \href{https://doi.org/10.1088/1475-7516/2018/08/048}{\emph{JCAP} {\bfseries
  08} (2018) 048} [\href{https://arxiv.org/abs/1806.09553}{{\ttfamily
  1806.09553}}].

\bibitem{Noorbala:2018zlv}
M.~Noorbala, V.~Vennin, H.~Assadullahi, H.~Firouzjahi and D.~Wands,
  \emph{{Tunneling in Stochastic Inflation}},
  \href{https://doi.org/10.1088/1475-7516/2018/09/032}{\emph{JCAP} {\bfseries
  09} (2018) 032} [\href{https://arxiv.org/abs/1806.09634}{{\ttfamily
  1806.09634}}].

\bibitem{Firouzjahi:2018vet}
H.~Firouzjahi, A.~Nassiri-Rad and M.~Noorbala, \emph{{Stochastic Ultra Slow
  Roll Inflation}},
  \href{https://doi.org/10.1088/1475-7516/2019/01/040}{\emph{JCAP} {\bfseries
  01} (2019) 040} [\href{https://arxiv.org/abs/1811.02175}{{\ttfamily
  1811.02175}}].

\bibitem{Biagetti:2018pjj}
M.~Biagetti, G.~Franciolini, A.~Kehagias and A.~Riotto, \emph{{Primordial Black
  Holes from Inflation and Quantum Diffusion}},
  \href{https://doi.org/10.1088/1475-7516/2018/07/032}{\emph{JCAP} {\bfseries
  07} (2018) 032} [\href{https://arxiv.org/abs/1804.07124}{{\ttfamily
  1804.07124}}].

\bibitem{Noorbala:2019kdd}
M.~Noorbala and H.~Firouzjahi, \emph{{Boundary crossing in stochastic inflation
  with a critical number of fields}},
  \href{https://doi.org/10.1103/PhysRevD.100.083510}{\emph{Phys. Rev. D}
  {\bfseries 100} (2019) 083510}
  [\href{https://arxiv.org/abs/1907.13149}{{\ttfamily 1907.13149}}].

\bibitem{Pattison:2019hef}
C.~Pattison, V.~Vennin, H.~Assadullahi and D.~Wands, \emph{{Stochastic
  inflation beyond slow roll}},
  \href{https://doi.org/10.1088/1475-7516/2019/07/031}{\emph{JCAP} {\bfseries
  07} (2019) 031} [\href{https://arxiv.org/abs/1905.06300}{{\ttfamily
  1905.06300}}].

\bibitem{Talebian:2019opf}
A.~Talebian, A.~Nassiri-Rad and H.~Firouzjahi, \emph{{Stochastic Effects in
  Anisotropic Inflation}},
  \href{https://doi.org/10.1103/PhysRevD.101.023524}{\emph{Phys. Rev. D}
  {\bfseries 101} (2020) 023524}
  [\href{https://arxiv.org/abs/1909.12773}{{\ttfamily 1909.12773}}].

\bibitem{Ezquiaga:2019ftu}
J.M.~Ezquiaga, J.~Garc\'\i{}a-Bellido and V.~Vennin, \emph{{The exponential
  tail of inflationary fluctuations: consequences for primordial black holes}},
  \href{https://doi.org/10.1088/1475-7516/2020/03/029}{\emph{JCAP} {\bfseries
  03} (2020) 029} [\href{https://arxiv.org/abs/1912.05399}{{\ttfamily
  1912.05399}}].

\bibitem{Firouzjahi:2020jrj}
H.~Firouzjahi, A.~Nassiri-Rad and M.~Noorbala, \emph{{Stochastic nonattractor
  inflation}}, \href{https://doi.org/10.1103/PhysRevD.102.123504}{\emph{Phys.
  Rev. D} {\bfseries 102} (2020) 123504}
  [\href{https://arxiv.org/abs/2009.04680}{{\ttfamily 2009.04680}}].

\bibitem{Ando:2020fjm}
K.~Ando and V.~Vennin, \emph{{Power spectrum in stochastic inflation}},
  \href{https://doi.org/10.1088/1475-7516/2021/04/057}{\emph{JCAP} {\bfseries
  04} (2021) 057} [\href{https://arxiv.org/abs/2012.02031}{{\ttfamily
  2012.02031}}].

\bibitem{Ballesteros:2020sre}
G.~Ballesteros, J.~Rey, M.~Taoso and A.~Urbano, \emph{{Stochastic inflationary
  dynamics beyond slow-roll and consequences for primordial black hole
  formation}}, \href{https://doi.org/10.1088/1475-7516/2020/08/043}{\emph{JCAP}
  {\bfseries 08} (2020) 043}
  [\href{https://arxiv.org/abs/2006.14597}{{\ttfamily 2006.14597}}].

\bibitem{Pattison:2021oen}
C.~Pattison, V.~Vennin, D.~Wands and H.~Assadullahi, \emph{{Ultra-slow-roll
  inflation with quantum diffusion}},
  \href{https://doi.org/10.1088/1475-7516/2021/04/080}{\emph{JCAP} {\bfseries
  04} (2021) 080} [\href{https://arxiv.org/abs/2101.05741}{{\ttfamily
  2101.05741}}].

\bibitem{Figueroa:2020jkf}
D.G.~Figueroa, S.~Raatikainen, S.~Rasanen and E.~Tomberg, \emph{{Non-Gaussian
  Tail of the Curvature Perturbation in Stochastic Ultraslow-Roll Inflation:
  Implications for Primordial Black Hole Production}},
  \href{https://doi.org/10.1103/PhysRevLett.127.101302}{\emph{Phys. Rev. Lett.}
  {\bfseries 127} (2021) 101302}
  [\href{https://arxiv.org/abs/2012.06551}{{\ttfamily 2012.06551}}].

\bibitem{Figueroa:2021zah}
D.G.~Figueroa, S.~Raatikainen, S.~Rasanen and E.~Tomberg, \emph{{Implications
  of stochastic effects for primordial black hole production in ultra-slow-roll
  inflation}}, \href{https://doi.org/10.1088/1475-7516/2022/05/027}{\emph{JCAP}
  {\bfseries 05} (2022) 027}
  [\href{https://arxiv.org/abs/2111.07437}{{\ttfamily 2111.07437}}].

\bibitem{Ahmadi:2022lsm}
N.~Ahmadi, M.~Noorbala, N.~Feyzabadi, F.~Eghbalpoor and Z.~Ahmadi,
  \emph{{Quantum diffusion in sharp transition to non-slow-roll phase}},
  \href{https://doi.org/10.1088/1475-7516/2022/08/078}{\emph{JCAP} {\bfseries
  08} (2022) 078} [\href{https://arxiv.org/abs/2207.10578}{{\ttfamily
  2207.10578}}].

\bibitem{Animali:2022otk}
C.~Animali and V.~Vennin, \emph{{Primordial black holes from stochastic
  tunnelling}},
  \href{https://doi.org/10.1088/1475-7516/2023/02/043}{\emph{JCAP} {\bfseries
  02} (2023) 043} [\href{https://arxiv.org/abs/2210.03812}{{\ttfamily
  2210.03812}}].

\bibitem{Talebian:2022jkb}
A.~Talebian, A.~Nassiri-Rad and H.~Firouzjahi, \emph{{Stochastic effects in
  axion inflation and primordial black hole formation}},
  \href{https://doi.org/10.1103/PhysRevD.105.103516}{\emph{Phys. Rev. D}
  {\bfseries 105} (2022) 103516}
  [\href{https://arxiv.org/abs/2202.02062}{{\ttfamily 2202.02062}}].

\bibitem{Nassiri-Rad:2022azj}
A.~Nassiri-Rad, K.~Asadi and H.~Firouzjahi, \emph{{Inflation with stochastic
  boundary}}, \href{https://doi.org/10.1103/PhysRevD.106.123528}{\emph{Phys.
  Rev. D} {\bfseries 106} (2022) 123528}
  [\href{https://arxiv.org/abs/2208.08229}{{\ttfamily 2208.08229}}].

\bibitem{Asadi:2023flu}
K.~Asadi, A.~Nassiri-Rad and H.~Firouzjahi, \emph{{Stochastic multiple fields
  inflation: Diffusion dominated regime}},
  \href{https://doi.org/10.1103/PhysRevD.108.123537}{\emph{Phys. Rev. D}
  {\bfseries 108} (2023) 123537}
  [\href{https://arxiv.org/abs/2304.00577}{{\ttfamily 2304.00577}}].

\bibitem{Tomberg:2023kli}
E.~Tomberg, \emph{{Stochastic constant-roll inflation and primordial black
  holes}}, \href{https://doi.org/10.1103/PhysRevD.108.043502}{\emph{Phys. Rev.
  D} {\bfseries 108} (2023) 043502}
  [\href{https://arxiv.org/abs/2304.10903}{{\ttfamily 2304.10903}}].

\bibitem{Mishra:2023lhe}
S.S.~Mishra, E.J.~Copeland and A.M.~Green, \emph{{Primordial black holes and
  stochastic inflation beyond slow roll. Part I. Noise matrix elements}},
  \href{https://doi.org/10.1088/1475-7516/2023/09/005}{\emph{JCAP} {\bfseries
  09} (2023) 005} [\href{https://arxiv.org/abs/2303.17375}{{\ttfamily
  2303.17375}}].

\bibitem{Tsamis:2003px}
N.C.~Tsamis and R.P.~Woodard, \emph{{Improved estimates of cosmological
  perturbations}},
  \href{https://doi.org/10.1103/PhysRevD.69.084005}{\emph{Phys. Rev. D}
  {\bfseries 69} (2004) 084005}
  [\href{https://arxiv.org/abs/astro-ph/0307463}{{\ttfamily
  astro-ph/0307463}}].

\end{thebibliography}\endgroup
\bibliographystyle{JHEP}
	
\end{document}